\documentclass[aps,pra,twocolumn,floatfix,superscriptaddress,showpacs,showkeys]{revtex4-1}
\usepackage{amsfonts}
\usepackage{amsmath}
\usepackage{amssymb}
\usepackage{amssymb}
\usepackage{graphicx}
\usepackage[english]{babel}
\usepackage{bm}% bold math
\begin{document}

\title{Fluorescence Excitation by Enhanced Plasmon Upconversion \\
under Continuous Wave Illumination}

\author{Mehmet Emre Tasgin}            
\affiliation{Institute of Nuclear Sciences, Hacettepe University, Beytepe, 06800 Ankara, Turkey}

\author{Ildar Salakhutdinov}
\affiliation{Center for Solar Energy Research and Applications (G\"{U}NAM), Middle East Technical University, Dumlup{\i}nar Blvd. 1, 06800, Ankara, Turkey}
\affiliation{Physics Department, Middle East Technical University, Dumlup{\i}nar Blvd. 1, 06800 Ankara, Turkey}

\author{Dania Kendziora}
\affiliation{DFG-Centre for Functional Nanostructures, Karlsruhe Institute of Technology, Wolfgang Gaede Str. 1, 76131 Karlsruhe, Germany}

\author{Musa Kurtulus Abak}
\affiliation{Center for Solar Energy Research and Applications (G\"{U}NAM), Middle East Technical University, Dumlup{\i}nar Blvd. 1, 06800, Ankara, Turkey}
\affiliation{Micro and Nanotechnology Program, Middle East Technical University, Dumlup{\i}nar Blvd. 1, 06800 Ankara, Turkey}

\author{Deniz Turkpence}
\affiliation{Department of Physics, Ko\c{c} University, \.{I}stanbul, Sar{\i}yer 34450, Turkey}

\author{Luca Piantanida}
\affiliation{CNR-IOM, Area Science Park, 34149 Trieste, Italy}
\affiliation{Graduate School of Nanotechnology, Department of Physics, University of Trieste, 34127 Trieste, Italy}

\author{Ljiljana Fruk}
\affiliation{DFG-Centre for Functional Nanostructures, Karlsruhe Institute of Technology, Wolfgang Gaede Str. 1, 76131 Karlsruhe, Germany}
\affiliation{Department of Chemical Engineering and Biotechnology, University of Cambridge, Pembroke Street, Cambridge CB2 3RA, United Kingdom}

\author{Marco Lazzarino}
\affiliation{CNR-IOM, Area Science Park, 34149 Trieste, Italy}
\affiliation{CBM S.c.r.l., Area Science Park, Basovizza, 34149 Trieste, Italy}

\author{Alpan Bek}
\email{bek@metu.edu.tr}
\affiliation{Center for Solar Energy Research and Applications (G\"{U}NAM), Middle East Technical University, Dumlup{\i}nar Blvd. 1, 06800, Ankara, Turkey}
\affiliation{Physics Department, Middle East Technical University, Dumlup{\i}nar Blvd. 1, 06800 Ankara, Turkey}
\affiliation{Micro and Nanotechnology Program, Middle East Technical University, Dumlup{\i}nar Blvd. 1, 06800 Ankara, Turkey}

\date{\today}

\begin{abstract}
We demonstrate effective background-free continuous wave nonlinear optical excitation of molecules that are sandwiched between asymmetrically constructed plasmonic gold nanoparticle clusters. We observe that near infrared photons are converted to visible photons through efficient plasmonic second harmonic generation. Our theoretical model and simulations demonstrate that Fano resonances may be responsible for being able to observe nonlinear conversion using a continuous wave light source. We show that nonlinearity enhancement of plasmonic nanostructures via coupled quantum mechanical oscillators such as molecules can be several orders larger as compared to their classical counterparts. 
\end{abstract}

%\pacs{xx}
\keywords{plasmons, second harmonic generation, Fano resonances, enhancement}

\maketitle
%%%%%%%%%%%%%%%%%%%%%%%%%%%%%%%%%%%%%%%%%%%%%%%%%%%%%%%%%%%%%%%%%%%%%%%%%%%%%%%%%%%%%%%%%%%%%%%%%%%%
%%%%%%%%%%%%%%%%%%%%%%%%%%%%%%%%%%%%%%%%%%%%%%%%%%%%%%%%%%%%%%%%%%%%%%%%%%%%%%%%%%%%%%%%%%%%%%%%%%%%
\section{Introduction}

Finding a possibility to control the activation of a single molecule is interesting since it paves the way to realize logic devices as small as molecules. This can be achieved by carefully designed devices which can interface a single molecule and interact with it at the same size scale. Such devices can be of electrical, magnetic or optical origin~\cite{Browne_2009} and may induce changes in the properties of a single or few molecules which can act as switching registers for data processing or data storage at the nanometer scale. Electrical devices at the single molecule size level include break-junctions~\cite{Xiang_2012}, scanning tunnelling microscope tip junctions~\cite{Tao_2003},  and molecular landers~\cite{Gourdon_1998}. Optical devices that can interface a single molecule among an ensemble of many molecules are yet lacking. The main difficulty lies in the diffraction limit of practically $\sim$ 150-180 nm focal size for a wavelength of 600 nm light. Compared to a few nm sizes of individual molecules, unambiguous activation of single molecules by means of absorption of light can only be possible for a device prepared from a sparse ensemble of molecules. 

Thanks to advancement in near-field optical instrumentation technology this main difficulty can be circumvented as the near-field light intensities typically have a very strong nonlinear dependence on distance (from $\sim r^{3}$ to $\sim r^{6}$). Strongly confined, highly intense fields of light called "hot spots", can be achieved by engineering the nanoscale environment of a single molecule. By utilizing plasmonic resonances of nanoscale metal structures, such spots can be arranged to occupy from a few down to a single molecule. Nevertheless in these schemes, a broad far-field background component of the same activation wavelength is inevitably superposed on to the near-field of the nanoparticles. This imposes limitations on single molecule activation experiments to be performed with high precision. When linear excitation schemes are employed, even a weak linear optical background signal as low as a single photon can give rise to activation of a single molecule as long as the excitation signal is within the absorption / activation spectrum of the molecule.

In an attempt to overcome the linear background problem, a nonlinear optical scheme namely SHG~\cite{Bautista_2012} can be used to address a single molecule. One approach to activate a single molecule is to use optical up-conversion of two simultaneously incident near-infrared (NIR) photons to create a single visible photon by the (localized) plasmonic activity of a metal nanoparticle cluster. It was shown earlier that asymmetric metal nanostructures are important for achieving second-order nonlinear optical response from plasmonic activity~\cite{Kauranen_2012,Girling}. An efficient second harmonic (SH) active plasmonic metal nanostructure can be constructed by composing a cluster of colloidal metal nanoparticles with two different sizes~\cite{Palomba_2009}. It is of essence to break the centro-symmetry of the cluster in order to have a nonvanishing second order susceptibility, $\chi^{(2)}$~\cite{Neacsu_2005,Zayats_overlap}. The strategy in this approach is to minimize the volume of nonlinear photoactivation by selectively activating the single or probably a few molecules which is/are located in the vicinity of the gap between the nanoparticles of such an asymmetric cluster. The photoactivation of a small number of molecules can be monitored by selecting the sandwiched molecules to possess a linear fluorescent nature and thereby absorb light only at the SH frequency  and not possess two-photon fluorescence cross-section at the irradiated frequency. 

In the absorption spectrum of hybrid structures that are made of coupled classical and quantum oscillators of charge –--such as a single molecule (or a quantum dot, etc.) attached to a plasmonic gold nanoparticle (AuNP)--– path interference effects similar to Fano resonances are observed~\cite{Manjavacas_2011,Artuso_2008,Waks_2010,Weis_2011}. The two absorption paths for the classical oscillator, induced by the hybridization with the quantum oscillator, can interfere destructively yielding a cancellation of the polarization on the plasmonic response for some frequencies~\cite{Tasgin_2010,Alzar}. Fano resonances also show up in the linear response of metal nanoparticles (MNPs) when a driven plasmon mode is coupled to a long-lived dark (usually quadrupole plasmon) mode~\cite{SoukoulisPRL2009}. Dark plasmon mode plays the role of the quantum object with a sharper spectrum.

Fano resonances also modify the nonlinear response of plasmonic nanoparticles. Enhancements in frequency conversion due to Fano resonances of interacting classical nanoparticles have already been studied numerically~\cite{Butet_2012} and observed experimentally~\cite{Thyagarajan_2013,Berthelot_2012,Walsh_2013}. In a recent experimental study~\cite{Thyagarajan_2013} an enhancement factor of 25 is reported. A systematic study of multiple Fano resonances, of full-classical nature, for the second harmonic generation (SHG) process is also performed in Ref.~\cite{MultiFano2016}. The observed enhancement in the nonlinear process can be explained simply by the coupling of the SH converter MNP to a long-lived dark plasmon mode~\cite{YildizJOpt2015}. The term induced by the dark mode cancels the nonresonant (i.e. $\Omega_2-2\omega$) in the denominator of the nonlinear response. This brings the conversion to the resonance. We note that, this enhancement is to be multiplied by the enhancement due to the field localization~\cite{localization_enhanced_SHG,surface_enhanced_SHG2}.

Recent studies~\cite{Turkpence_2014,MahiSingh2013,Tasgin_2014} reveal that Fano resonances of coupled plasmonic-quantum systems can also enhance the SHG and four-wave mixing~\cite{TasginPRB2016} processes. In our theoretical and numerical studies  using 3D simulations with retardation effects~\cite{Turkpence_2014,TasginPRB2016,YildizJOpt2015,Tasgin_2014},  we have demonstrated that path interference schemes can be adopted to gain control over the nonlinear frequency conversion processes emerging in plasmonic resonators. By appropriately choosing the level spacing of the quantum oscillator (i.e. choosing the right molecule), one can either enhance or suppress the frequency conversion process by several orders of magnitude (see Sec.~\ref{sec:path_inter}).

%
%In our recent theoretical and numerical studies  using 3D simulations with retardation effects~\cite{Turkpence_2014,TasginPRB2016,YildizJOpt2015,Tasgin_2014}, we have demonstrated that path interference schemes can be adopted to gain control over the nonlinear frequency conversion processes emerging in plasmonic resonators. By appropriately choosing the level spacing of the quantum oscillator (i.e. choosing the right molecule), one can either enhance or suppress the frequency conversion process by several orders of magnitude.  An enhancement in the nonlinear response emerges whenever the path interference is arranged to cancel the nonresonant terms, thereby bringing the conversion to resonance. This enhancement/suppression is to be multiplied by the enhancement due to the field localization~\cite{localization_enhanced_SHG,surface_enhanced_SHG2}. Enhancements in frequency conversion due to Fano resonances of interacting classical nanoparticles have already been studied numerically~\cite{Butet_2012} and observed experimentally~\cite{Thyagarajan_2013,Berthelot_2012,Walsh_2013}. In a recent experimental study~\cite{Thyagarajan_2013} an enhancement factor of 25 is reported. A systematic study of multiple Fano resonances, of full-classical nature, for the SHG process is also performed in Ref.~\cite{MultiFano2016}.

{\it In this paper}, we show that fluorescent molecules –--with absorption bands in the visible part of the spectrum--– can be brought to fluoresce under continuous wave (cw) near infrared (NIR) excitation (1064 nm). We prepare two different  clusters, (i) 12 nm sized AuNPs decorated with enhanced yellow fluorescent protein (EYFP) molecules [Fig.~\ref{Figure 1} (a)-(b)] and (ii) bare AuNPs of sizes between 50 - 120 nm. We observe that upon cw excitation with NIR laser  these two clusters do not express SHG or fluorescence. However when the two clusters are brought together [Fig. \ref{Figure 1} (c)-(d)], we observe visible fluorescence from the molecules.

We argue that fluorescence is realized by absorption of the up-converted plasmon excitation (532 nm) from the AuNP clusters. We consider the presence of the SHG process, since the converted wavelength lies within the linear absorption spectrum of the molecule (Fig.~\ref{Figure 2}). We exclude the possibility of the two-photon conversion by the molecule (see Sec.~\ref{sec:upconversion}). This is because the molecule does not exhibit two-photon absorption at the excitation wavelength of 1064 nm (see Fig.~\ref{Figure 2}). We can also understand, from the 2-photon fluorescence spectrum of the EYFP-AuNP (in Fig.~\ref{Figure 4}), that two-photon absorption spectrum of the molecule is not modified considerably when they are coupled to the AuNPs.

We provide an explanation for being able to observe the high-frequency signal only when the two clusters are brought together by numerically solving Maxwell's equations in 3D~\cite{Hohenester_2012}] and introducing a model for the SHG enhancement due to the path interference. We demonstrate that when the 12nm AuNP (decorated with EYFP molecules) is placed between two larger size AuNPs (Fig.~\ref{Figure 5}), 12 nm AuNP supports a quadrupole plasmon mode (Fig.~\ref{Figure 6}) at the upconverted frequency region. Presence of this quadrupole mode is very important for the emergence of the frequency conversion. Because the overlap integral for the SHG process possesses symmetry selection rules~\cite{Neacsu_2005,Zayats_overlap}, see Eq.~(\ref{chi2}) and discussion below. 1064 nm excitation produces dipole-like plasmon oscillations (Fig.~\ref{Figure 7}). Selection rules dictate that SH oscillations can emerge into a mode with an even spatial profile (e.g. a quadrupole mode).

We also demonstrate that presence of the molecules, interacting with this high-energy plasmon mode can enhance the nonlinear response (SH conversion) about $\sim$1000 times via path interference effects (Fig.~\ref{Figure 8}). Still, observation of the SH oscillations in the far-field would not be possible if the molecules would not absorb and fluoresce the upconverted plasmon oscillations. This is because the quadrupole modes couple to the far-field weakly ---see the matrix element (20) in Ref.~\cite{2nd_quantized}.

We consider that (i) emergence of the quadrupole-like mode in the 12 nm AuNP via hybridization with the larger AuNPs, (ii) spectral shift of the plasmon mode (supporting 1064 nm oscillations) to infrared, (iii) serious enhancement factor due to path interference induced by the molecules, and (iv) communication of the molecules with the far-field via fluorescence, make the observation of the high-frequency (529 nm) signal possible.

The experimental details on preparation of such a sample and excitation measurement results are provided in Sec.~\ref{exp}. The 3D simulations of the hybridized AuNPs, emergence of the quadrupole-like plasmon mode, origin of the selection rules for SH process, a model describing the enhancing/suppressing effects of the path interference on the nonlinear response, and a discussion on the control of these effects are given in Sec.~\ref{theo}. In Sec.~\ref{sec:upconversion}, we discuss other possibilities for the mechanism of the observed up-converted signal in the experiment. Finally, in Sec.~ \ref{con} we summarize our experimental and theoretical results. 

\section{Experiment} {\label{exp}}
\subsection{Experimental Methods}

AuNPs were synthesized according to literature~\cite{Storhoff_1998}. Glassware was cleaned with aqua regia, thoroughly washed with water and dried at 70$^o$C. In a three-neck flask 25 mL chloroauric acid (1 mM, 25 μmol, 1 eq.) was heated under reflux to 100$^o$C. After addition of 2.50 mL trisodium citrate (38.8 mM, 97 $\mu$mol, 3.88 eq.) the solution was stirred for 1 h and cooled to room temperature. Resulting AuNP were analyzed with UV-Vis spectroscopy and transmission electron microscopy (TEM). The diameter was determined to 11.7 $\pm$ 0.96 nm. Concentration of AuNP was estimated using a literature known extinction coefficient of $\epsilon_{450nm}=6.15\times 10^7$ M$^{-1}$cm$^{-1}$ for citrate coated AuNP~\cite{Haiss_2007}.

Larger, nondecorated AuNPs of 50-120 nm size were synthesized following a revised version of the well-known Turkevich method~\cite{Turkevich_1951,Enüstün_1963,Ojea-Jiménez_2010,Kimling_2006}. Sodium citrate (0.86 mM) was dissolved in 100 mL of milliQ water and the solution was heated up to 100$^o$C without refluxing system. At the boiling point a solution of HAuCl$_{4}$ : 3 H$_{2}$O (1 mL, 0.04 mM) was added and the mixture was left boiling for 5 min before cooling down to room temperature (RT). 

DNA-modified AuNP were formed by ligand exchange with thiolated oligonucleotides as described in an earlier publication~\cite{Hazarika_2006}. Briefly, 100 $\mu$L thiolated oligonucleotide (100 $\mu$M, 10 nmol, 1 eq.) were reduced with 60 $\mu$L dithiothreitol (DTT, 1 M, 60 $\mu$mol, 6000 eq.) over night at 37$^o$C. The reaction mixture was purified with gel filtration using NAP5 und NAP10 columns (GE Healthcare). To the resulting solution 1 mL citrate-stabilised 12 nm AuNP (30 nM, 30 pmol, 0.003 eq.) were added and incubated overnight at RT. After addition of 2 mL TETBS300 (20 mM Tris, 300 mM NaCl, 5 mM EDTA, 0.05$\%$ Tween-20, pH 7.5) the solution was kept overnight at RT. To remove unbound oligonucleotides AuNP were centrifuged 30 min at 13200 rpm, the supernatant was removed, the residue was dissolved in TETBS 300 and the step repeated twice. 

EYFP was modified with DNA as described in an earlier publication~\cite{Kukolka_2004}. Briefly, 200 $\mu$L EYFP (200 $\mu$M, 40 nmol, 1 eq.) were reduced with 60 $\mu$L DTT (1 M, 60 $\mu$mol, 1500 eq.) for 2 h at 23$^o$C. Simultaneously 100 $\mu$L amino-cDNA (100 $\mu$M, 10 nmol, 1 eq.) was modified with 100 $\mu$L sSMCC (46 mM, 4.60 $\mu$mol, 460 eq.) for 2 h at 23$^o$C to yield maleimide-cDNA. After purification with NAP5 and NAP10 columns, both solutions were combined and incubated overnight at 4$^o$C. The reaction mixture was concentrated and the buffer changed to 20 mM Tris, pH 8.3 before purification with anion exchange chromatography using a MonoQ 5/50 (GE Healthcare). Signals containing desired product were identified using UV-Vis spectroscopy and native TBE-polyacrylamid gel electrophoresis. To form EYFP-modified AuNP 38.5 $\mu$L AuNP-DNA (5.20 nM, 0.20 pmol, 1 eq.) were incubated with 2.67 $\mu$L EYFP-cDNA (1.87 $\mu$M, 5 pmol, 25 eq.) at room temperature overnight~\cite{Niemeyer_2003}. The resulting conjugate was analyzed with agarose gel electrophoresis.

\begin{figure}[!t]
\begin{center}
\includegraphics[width=3.2in]{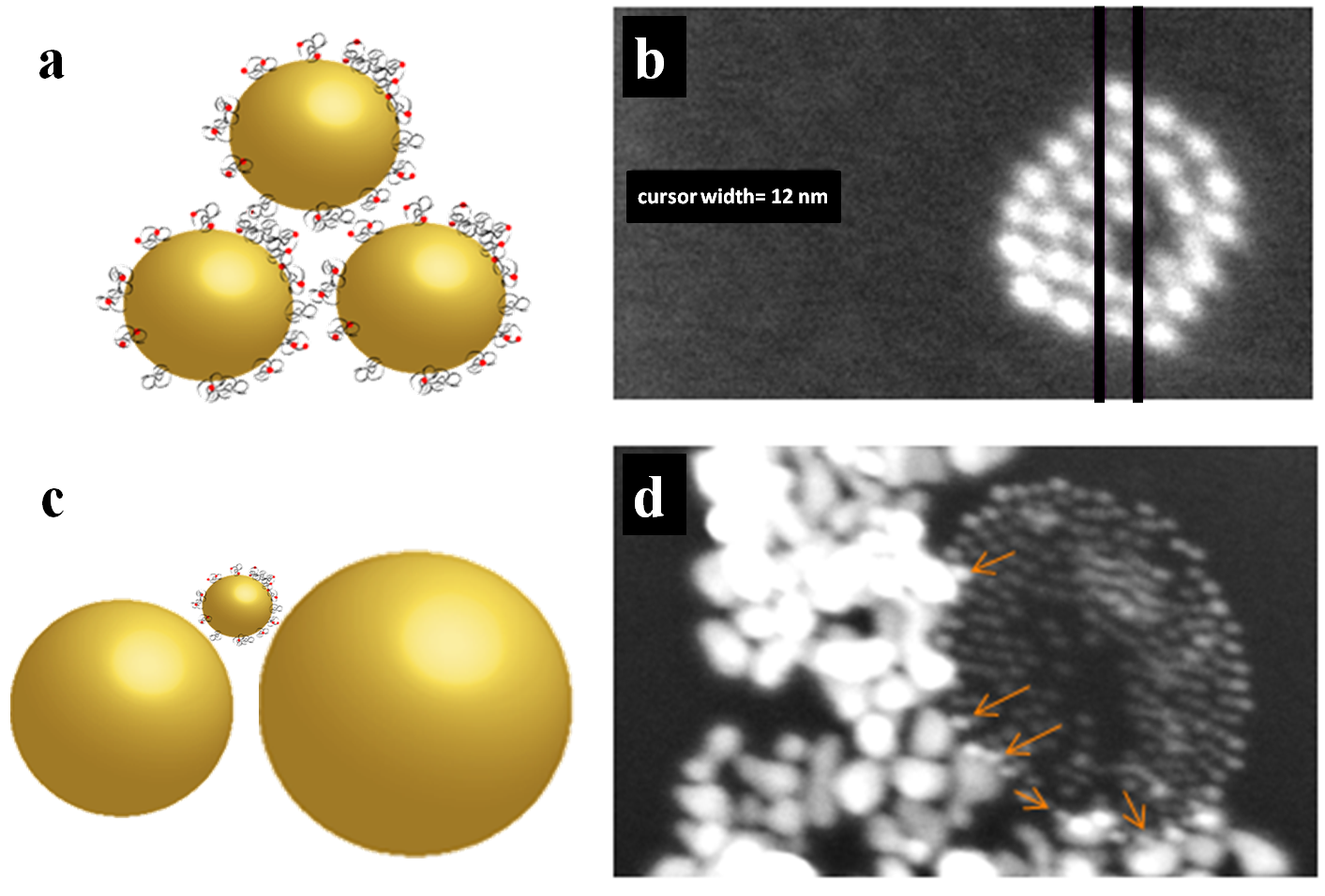}
\caption{\label{Figure 1} (color online) Symmetric and asymmetric AuNP clusters. (a) The EYFP decorated Au nanoparticle, (b) An SEM image of a symmetric cluster of such nanoparticles indicating the size of the Au nanoparticles of $\sim$12 nm, (c) The sketch of an asymmetric AuNP dimer with a single/few molecule/s in the gap, (d) An SEM image of a surface prepared using both types of nanoparticles displaying such asymmetric clusters as indicated by arrows.} 
\end{center}
\end{figure}

In our experiment we constructed a gold cluster complex using EYFP decorated AuNPs together with larger diameter bare AuNPs (Fig.~\ref{Figure 1}). The cluster structure was formed by mixing two droplets of solution containing the two different types of nanoparticles (12 nm and 50-120 nm) on the atomically flat (100) surface of a silicon wafer. Figures~\ref{Figure 1} (a)-(d) show a sketch of a single EYFP decorated AuNP, an SEM image of a cluster of such nanoparticles, a sketch of an asymmetric AuNP dimer with EYFPs in the vicinity of the gap, and an SEM image of such clusters, respectively. Au is chosen due to its matching plasmon resonance of 515-560 nm with the SH wavelength of 532 nm, belonging to the irradiation laser of 1064 nm.

\begin{figure}[!t]
\begin{center}
\includegraphics[width=3.2in]{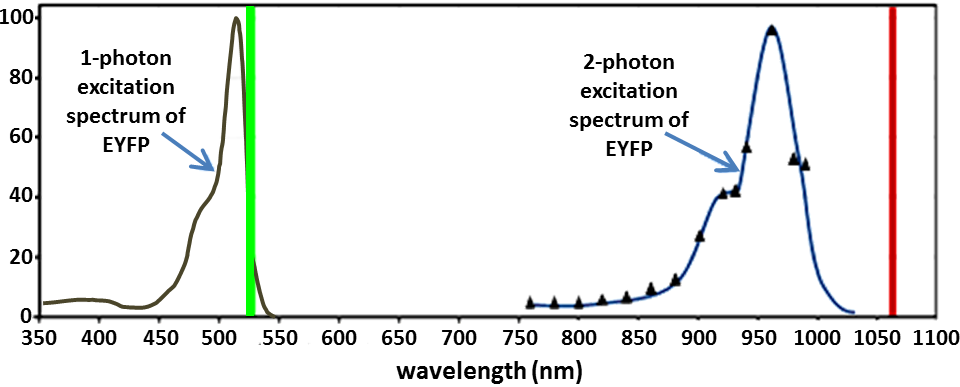}
\caption{\label{Figure 2} (color online) The normalized 1- and 2-photon excitation spectra of EYFP. The excitation wavelength of 1064 nm and its second harmonic (532 nm) are marked by vertical red and green lines, respectively. (adapted from \cite{Blab_2001})} 
\end{center}
\end{figure}

Our experiment is based on the idea that EYFP is sensitive to 1064 nm radiation neither linearly nor nonlinearly as can be seen from Fig. \ref{Figure 2}, since the linear and two-photon absorption bands lie between 430-550 nm and 850-1030 nm, respectively~\cite{Blab_2001}. It is of essential importance to our experiment that there exists a significant shift between twice the one-photon and the two-photon excitation band maxima of EYFP, so that direct two-photon excitation of EYFP by irradiation at 1064 nm is avoided. We experimentally confirmed that; and the details are provided in the next section. So that when irradiated with 1064 nm of light the EYFP molecule can generate a fluorescence signal only if 1064 nm light were up-converted to 532 nm by the asymmetric AuNP cluster. The linear emission band maximum of EYFP is located at 529 nm which is very close to SH signal from the laser.

As we can see from Fig.~\ref{Figure 1}(b), there are areas where smaller nanoparticles decorated by EYFP are clustered. We used an inverted microscope's (Zeiss model Axiovert 200) 63X (1.4 NA) objective to identify these areas. Areas with the highest concentration of EYFP were identified by scanning a 532 nm DPSS laser (Cobolt model Samba), and detecting the maxima of the fluorescence signal.

\begin{figure}[!t]
\begin{center}
\includegraphics[width=3.2in]{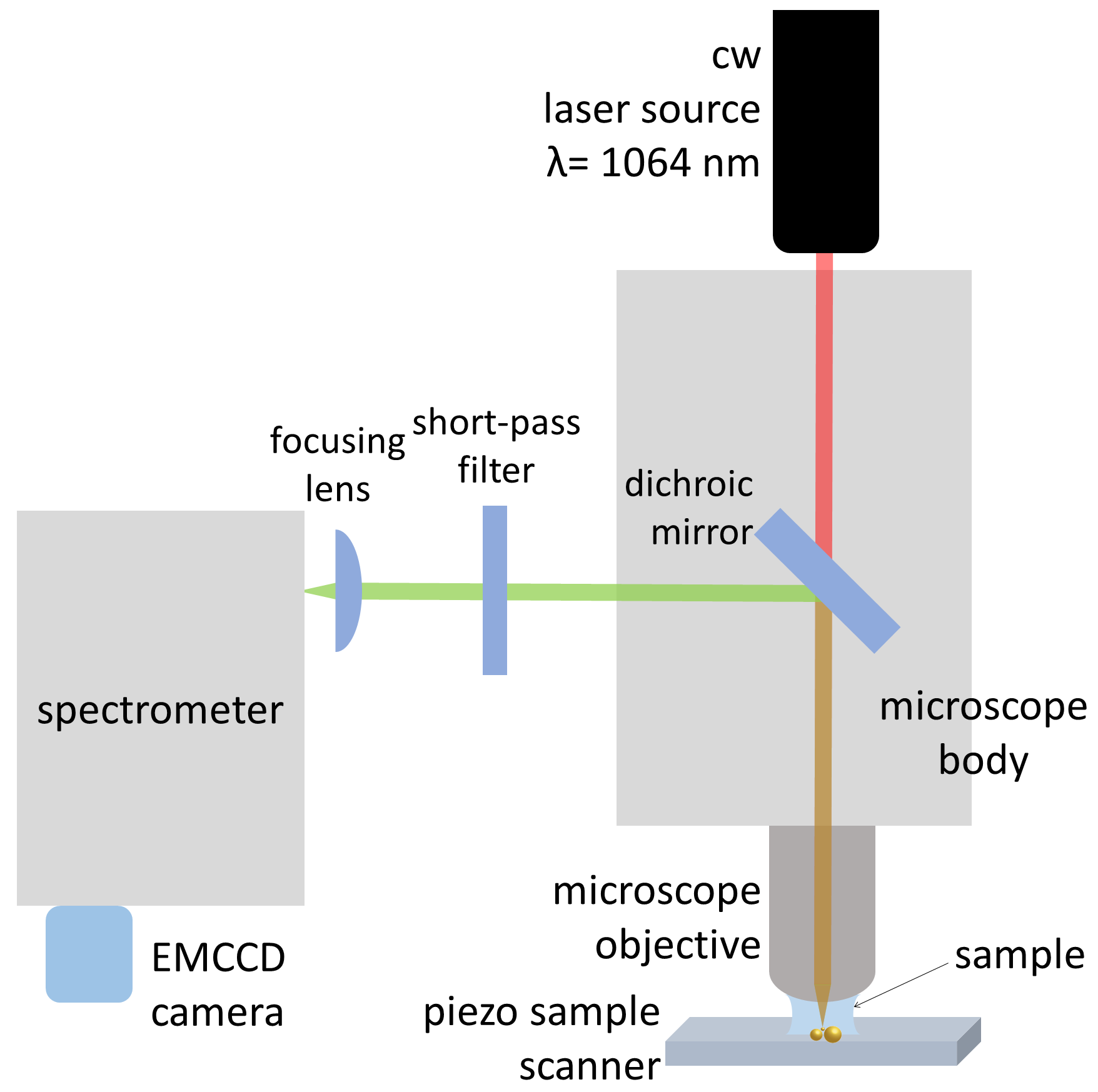}
\caption{\label{Figure 3} (color online) The experimental setup consists of a cw Nd:YAG laser with $\lambda $ = 1064 nm, a long-pass dichroic mirror,  a microscope objective, sample, a focusing lens, a short-pass filter (cut-off at 800 nm), a spectrometer with an  EMCCD camera.} 
\end{center}
\end{figure}

In Fig.~\ref{Figure 3}, a scheme of experimental setup is displayed. We used a cw Nd:YAG laser (Cobolt model Rumba) with wavelength of $\lambda$ = 1064 nm and 500 mW output power as excitation source. The NIR signal was delivered from the back port of microscope, resulting in an intensity of 40 MW/cm$^2$ on the sample. Resulting fluorescence from the clusters was registered by a spectrometer (Andor models Shamrock 750 spectrograph + Newton 971 EMCCD) in cooperation with a long-pass dichroic mirror cutting off the backscattered 1064 nm signal and reflecting a band of 532-750 nm. 

%%%%%%%%%%%%%%%%%%%%%%%%%%%%%%%%%%%%%%%%%%%%%%%%%%%%%%%%%%%%%%%%%%%%%%%%%%%%%%%%%%%%%%%%%%%%%%%%%%%%%%%%
%%%%%%%%%%%%%%%%%%%%%%%%%%%%%%%%%%%%%%%%%%%%%%%%%%%%%%%%%%%%%%%%%%%%%%%%%%%%%%%%%%%%%%%%%%%%%%%%%%%%%%%%
%%%%%%%%%%%%%%%%%%%%%%%%%%%%%%%%%%%%%%%%%%%%%%%%%%%%%%%%%%%%%%%%%%%%%%%%%%%%%%%%%%%%%%%%%%%%%%%%%%%%%%%%
\subsection{Experimental Results}

First step in our experiments involved the certification that the direct two-photon excitation of EYFP is avoided by 1064 nm laser irradiation. We have performed control measurements on the clusters of EYFP decorated 12 nm AuNPs in the absence of 50-120 nm AuNP cluster using the same excitation power as the experiments involving both clusters. The measurements yielded no two-photon fluorescence signal from 12 nm AuNP clusters hosting EYFP molecules. As a second control experiment, we tested SHG activity of the 50-120 nm AuNP clusters in the absence of 12 nm EYFP AuNP clusters, and similarly observed no SHG converted signal.

\begin{figure}[!t]
\begin{center}
\includegraphics[width=3.2in]{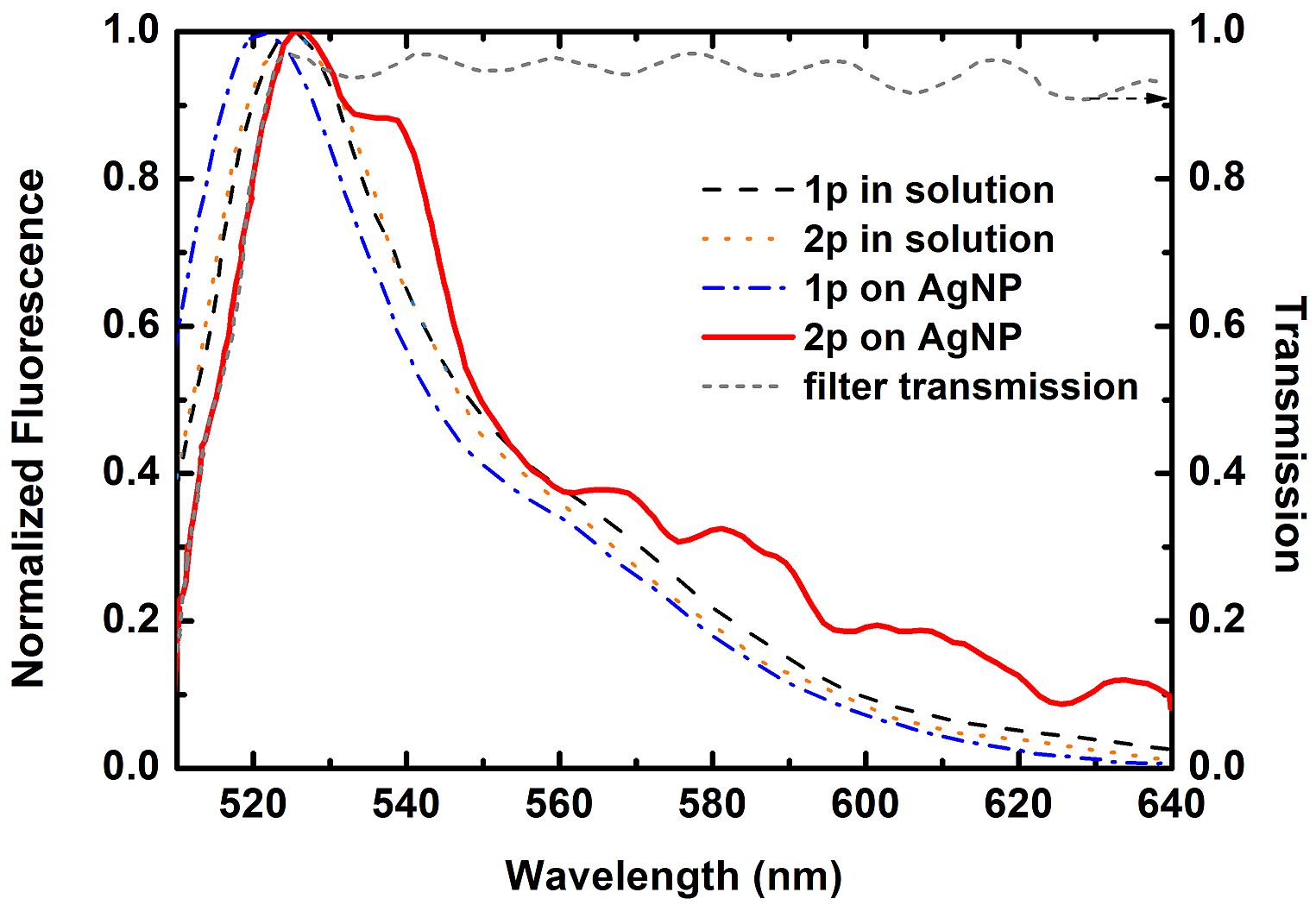}
\caption{\label{Figure 4} (color online) The fluorescence spectra of EYFP excited both by one-photon (black) and plasmon induced two-photon processes (red) obtained from the asymmetric clusters. 1-photon fluorescence is adopted from Ref.~\cite{fluorescenceEYFP} where EYFP molecules are contained in a solution. Note that peak of the 2-photon fluorescence with AuNPs (2p on AuNP) curve looks shifted due to the presence of the filter.  } 
\end{center}
\end{figure}

Fig.~\ref{Figure 4} presents the EYFP fluorescence data obtained from EYFP molecules that are sandwiched in-between plasmonic asymmetric structures. The red curve corresponds to fluorescence from such aggregates under 1064 nm illumination. The black curve is the fluorescence obtained by a combination of halogen lamp and a FITC filter set (Zeiss) for a comparison. The green curve is the transmission curve belonging to the dichroic mirror used in this experiment. As expected, the plasmon induced two-photon fluorescence was much weaker compared to the linear fluorescence, by an order of 10$^{-12}$. This is quite comprehensive since in two-photon excitation, the fluorescence signal originates from only a single or a few molecules sandwiched right in the gap of a AuNP cluster, whereas the linear fluorescence signal originates from all the EYFP molecules besides those whose fluorescence may be quenched by the AuNPs, yet quenching affects both the sandwiched and non-sandwiched molecules equally. A slight shift of fluorescence maximum from the linear emission band maximum of EYFP at 529 nm to $\sim$535 nm, and a low level modulation of the fluorescence band towards longer wavelengths can be observed. This is a consequence of the transmission properties of the dichroic mirror as the green curve in Fig.~\ref{Figure 4} suggests. The importance of asymmetry was tested by control experiments in which same size ($\sim$12 nm) nanoparticles were let to form symmetric AuNP clusters, from which a fluorescence signal upon NIR illumination was absent. This confirms that asymmetry of the plasmonic structure plays a crucial role for achieving second harmonic response from plasmonic structures.

Next, we demonstrate that the EYFP molecules are responsible for the enhancement of the SH production due to presence of Fano resonances, discussed in Sec.~\ref{sec:path_inter} in detail, as well as reporting the generated $2\omega$ localized surface plasmon(LSP) oscillations. We also discuss the assisting role of the hybridization in SH conversion process. Here $\omega$ represents the drive frequency corresponding to $\lambda=1064$ nm excitation wavelength.

\begin{figure}[!t]
\begin{center}
\includegraphics[width=3.2in]{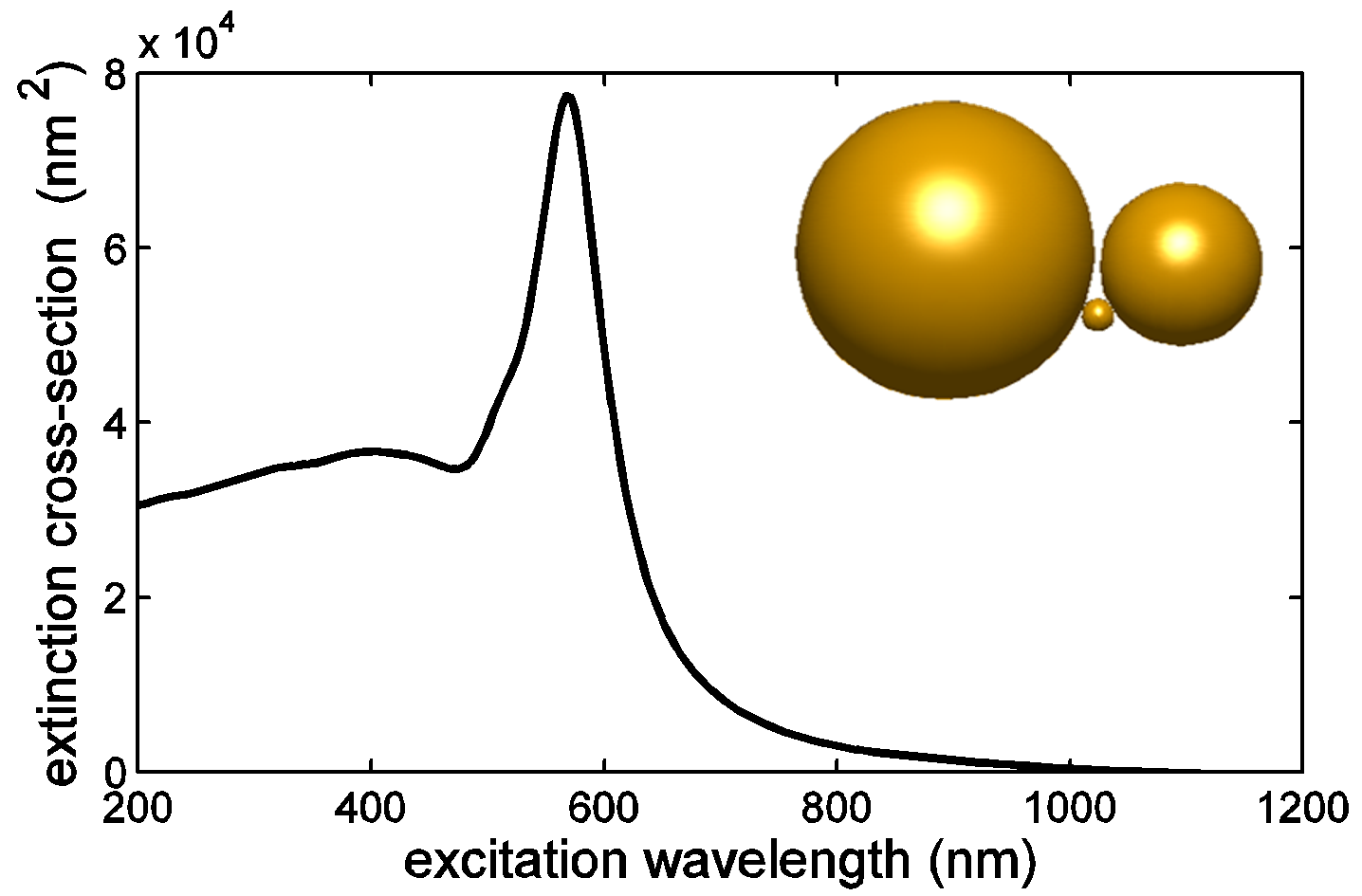}
\caption{\label{Figure 5} (color online) The extinction cross-section of three, of sizes 110nm 60nm and 12nm, hybridized AuNPs. The first LSP mode, mode 1 with resonance peak at $\lambda_1=580$ nm, has an extended tail at the drive wavelength of $\lambda=1064$ nm. The second LSP mode, mode 2, has resonance peak at $\lambda_2=400$ nm. Due to the strong localization of the plasmon polarization field, two LSP excitations in mode 1 oscillating at $\lambda=1064$ nm combine to generate a $\lambda/2=532$ nm oscillation in the mode 2. Even though, SHG at $\lambda/2=532$ nm is off-resonant with $\lambda_2$, insertion of a molecule with an absorption peak at $\lambda_{abs}=514$ nm enhances the SHG (see Fig.~\ref{Figure 8}) substantially. (The separation between the 110nm and 60nm AuNPs is 2nm. 12nm particle is placed $\sim$1nm distance to both AuNPs.)} 
\end{center}
\end{figure}

The extinction peak of a single 70 nm AuNP (the median size of 50-120 nm AuNPs) is at about $\lambda_1\cong 510$ nm which is quite far away from  $\lambda=1064$ nm, the wavelength of the excitation laser we use. This mode is scarcely affected by the diameter of the AuNP. When AuNPs are brought together, they hybridize to yield plasmon resonances at longer wavelengths~\cite{Nordlander_2004}. As an example, in Fig.~\ref{Figure 5}, we present the hybridized spectrum of three AuNPs at different sizes, 110 nm, 60 nm and 12 nm. This is one of the possible scenarios taking place at the interface, where a 12 nm AuNP (decorated with molecules) goes in between two larger AuNPs. The low-lying resonance shifts to 580 nm and its tail reaches longer wavelengths. The scattering cross-section grows from 0.44 nm$^2$ to 726 nm$^2$. This effect is proposed to be used for wide spectrum conversion in solar cells~\cite{Günendi_2013}, and gives rise to increased absorption of the incident radiation. Hybridization also assists SHG by shifting the second plasmon excitation peak from 270 nm, for single AuNP, to 400 nm in hybrid (aggregate) AuNPs.

In our experiment, in contrast to larger AuNPs, 12 nm AuNPs require smaller separations (e.g. $\sim $0.3 nm) to express a similar hybridization. This is due to the relatively small spatial extension of the plasmonic excitations. Moreover, the presence of molecules surrounding the 12 nm AuNPs inhibits the hybridization. Hence, one does not observe SHG from 12 nm AuNP aggregates alone due to the weakness of the hybridization, in addition to the centrosymmetric nature of such particles.

%%%%%%%%%%%%%%%%%%%%%%%%%%%%%%%%%%%%%%%%%%%%%%%%%%%%%%%%%%%%%%%%%%%%%%%%%%%%%%%%%%%%%%%%%%%%%%%%%%%%%%%%
%%%%%%%%%%%%%%%%%%%%%%%%%%%%%%%%%%%%%%%%%%%%%%%%%%%%%%%%%%%%%%%%%%%%%%%%%%%%%%%%%%%%%%%%%%%%%%%%%%%%%%%%
%%%%%%%%%%%%%%%%%%%%%%%%%%%%%%%%%%%%%%%%%%%%%%%%%%%%%%%%%%%%%%%%%%%%%%%%%%%%%%%%%%%%%%%%%%%%%%%%%%%%%%%%
\section{Theory and Simulations} {\label{theo}}

In two recent theoretical works~\cite{Turkpence_2014,Tasgin_2014} 3D boundary element method (MNPBEM~\cite{Hohenester_2012}) simulations demonstrate that second harmonic generation in plasmonic resonators can be enhanced by using the Fano resonances. In these simulations, the retardation effects are taken into account, where it is confirmed that the retardation effects do not play a significant role on the existence of SHG enhancement.

Nonlinear optical response is strongly modified upon introduction of a coupled quantum emitter which has a small decay rate. When the level spacing of the quantum emitter is resonant with the exact frequency of $2\omega$, SHG can be suppressed several orders of magnitude~\cite{Berthelot_2012}. Transparency induced by the Fano resonance, similar to linear case, does not allow the excitation of the second harmonic mode. In contrast, when quantum level spacing is chosen to be in the close proximity of $2\omega$ (e.g. $\omega_{eg}=2.04\omega$), SHG can be enhanced 2 orders of magnitude. Such an enhancement occurs since the path interference effects cancel nonresonant terms [see Eq.~(\ref{EQ8}) below] and bring the nonlinear conversion to resonance~\cite{Turkpence_2014}. In the following part, we discuss the enhancement mechanism within the framework of Fano resonance. From our simulations we argue that such an enhancement scheme is in charge of the observed highly enhanced SHG in our experiment.

A previously published k-space spectroscopy experiment~\cite{Grosse_2012} reveals that the mechanism of the SHG process in plasmonic resonators is as follows. The field is strongly localized in terms of LSP oscillations ($\omega$). Two of such LSP oscillations ($\omega$) combine to produce a single high frequency ($2\omega$) LSP. SH photons (at $2\omega$) originate from the radiative-decay of this LSP excitation~\cite{Bouhelier_2005,Note}. (Radiative decay is weak for even plasmon modes.) Therefore, the Hamiltonian for the process can be written as~\cite{Turkpence_2014}
\begin{equation}
H_{sh}=\hbar\chi^{(2)}\left( \hat{a}^{\dagger}_2\hat{a}_1\hat{a}_1+ \hat{a}^{\dagger}_1\hat{a}^{\dagger}_1\hat{a}_2         \right).\label{EQ1}
\end{equation}
Here, $\hat{a}_1$ is the driven LSP mode, and $\hat{a}_2$ is the LSP mode in which SH plasmon excites. $\chi^{(2)}$ is proportional to the overlap integral~\cite{Zayats_overlap}
\begin{equation}
\chi^{(2)} \propto \int d^3{\bf r} E_2^{*}({\bf r}) E_{1}^2({\bf r}) \rho({\bf r}) + C.c.
\label{chi2}
\end{equation}
which determines (also proportional to) the SH dielectric susceptibility. $E_{1}({\bf r})$ is the spatial profile of the plasmon mode driven by the CW laser,  of resonance $\omega_1$ or $\lambda_1$ below. $ E_2({\bf r})$ is the spatial profile of the higher energy plasmon mode, of resonance $\omega_2$ or $\lambda_2$ below, in which SH oscillations emerge. $\rho({\bf r})$ is the profile of the medium where SH conversion process takes place, that is the MNP(s). For simplicity, $\rho({\bf r})$ can be considered as a 3D $\theta({\bf r})$ step function. Such a standard overlap integral can be obtained after elimination of the operators responsible for other participated processes, e.g. inter-band transitions~\cite{Neacsu_2005,Senet_1996,Lohner_1999,Schaich_2000,Bachelier_2010,Liebsch_1988}. 

Eq.~(\ref{chi2}) possesses some symmetry rules for the presence of SHG~\cite{Neacsu_2005}. Lower energy plasmon mode $E_{1}({\bf r})$ has a dipolar nature in optical excitation wavelengths (in our experiment/simulations). Hence, if the metal nanoparticle --on which SHG takes place-- has the spherical symmetry, $E_{2}({\bf r})$ needs to be an even function, that is quadrupole profile. This is because, $E_{1}^2({\bf r})$ becomes an even function.  When SHG takes place in the hot-spot of two interacting MNPs, which has dipole-like character, again a mode with even spatial profile (about the gap) is required.

Quadrupole modes couple to the far-field very weakly~\cite{Pelton_book_2013}, see the matrix element (20) in Ref.~\cite{2nd_quantized}. Unlike the standard experiments with strong pulsed lasers, we use a CW laser for the pumping. When the $2\omega$ oscillations arise in a quadrupole mode, their radiative decay to the far-field may not be observed for relatively weak CW illumination. For this reason, we use EYFP as reporter molecule which is excited by interaction with the $\hat{a}_2$ LSP mode and fluoresce to far field. In our simulations, below, we observe that $E_2({\bf r})$ mode indeed coincides with a quadrupole profile on the surface of the 12 nm AuNP, see Fig.~\ref{Figure 6}.

\subsection{Quadrupole charge profile and nonzero $\mathbf{\chi^{(2)}}$}

As a sample configuration, we simulate the plasmon modes of a hybridized AuNPs system, depicted in Fig.~\ref{Figure 5}, with 3 particles of sizes 110 nm, 60 nm and 12 nm. In the experiment, 12 nm AuNP is decorated with the EYFP molecules. For this configuration we observe that  $\lambda_1=580$ nm and $\lambda_2=400$ nm correspond to resonances $a_1$ and $a_2$, ($\omega_1<\omega_2$), respectively. 1064 nm ($\omega$) radiation excites $\lambda_1=580$ nm LSP mode and SH plasmon is created in $\lambda_2=400$ nm LSP mode. We note that, SH plasmon cannot be created in the driven $\hat{a}_1$ LSP mode since the excitations are to be presented with the same operator $\hat{a}_1$ and energy conservation inhibits such a term.
 
\begin{figure}%[!t]
\begin{center}
\includegraphics[width=3in]{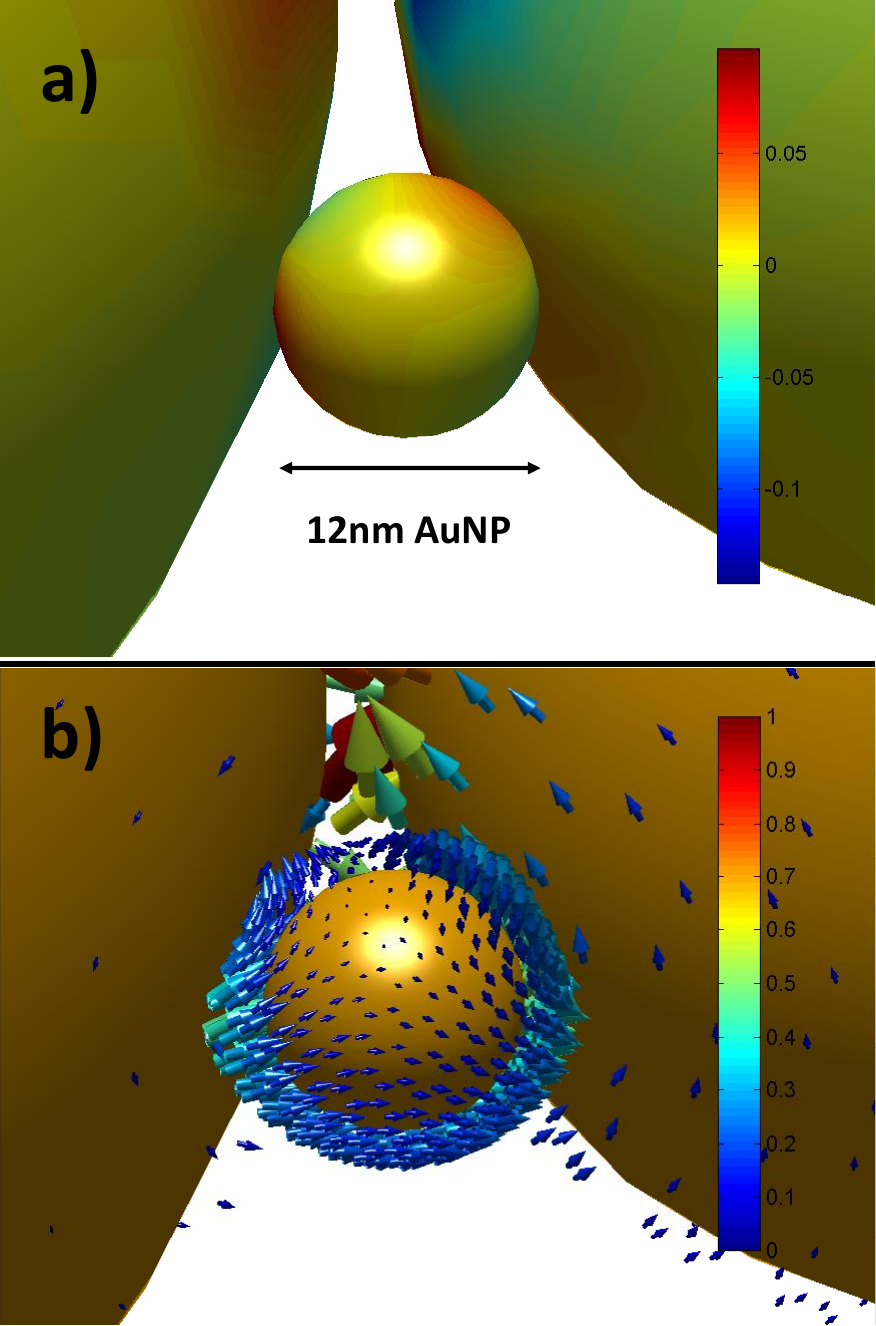}
\caption{\label{Figure 6} (color online) (a) Charge distribution of the eigen-mode, of eigen-frequency $\sim$400 nm, exhibits a quadrupole distribution on the 12 nm AuNP. The red and blue charge peaks, on the 12 nm AuNP, are almost equal with opposite sings. The simulation~\cite{Hohenester_2012} is performed for the configuration considered in Fig.~\ref{Figure 5}. Since the mode is quadrupole-like, that is an even function, the integral (\ref{chi2}) does not vanish when the system is excited with 1064 nm dipole-like (see Fig.~\ref{Figure 7}) plasmon modes. (b) Distribution of the electric field for the same eigen-mode. An EYFP molecule attached to the 12 nm AuNP particle interacts (stronger on the top of the 12 nm AuNP) with the 532 nm oscillations created in the quadrupole mode. This interaction leads to the Fano resonance with the $\lambda_2$ mode and results in the 2-3 orders of magnitude enhancement of the SH conversion process, see Eq. (\ref{EQ8}) and the text below. Units are arbitrary.   } 
\end{center}
\end{figure}

\begin{figure}%[!t]
\begin{center}
\includegraphics[width=3in]{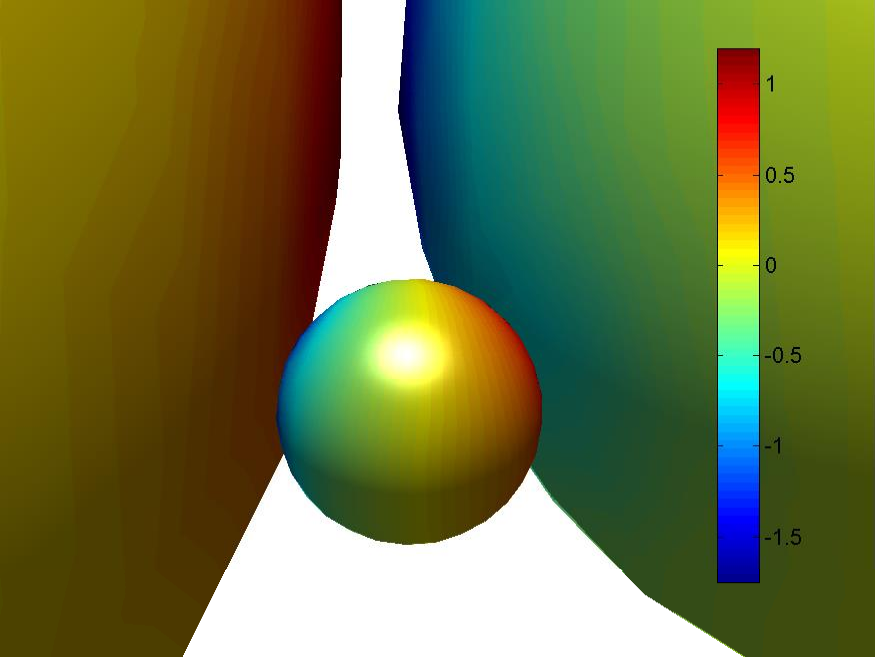}
\caption{\label{Figure 7} (color online) Charge distribution of the configuration considered in Fig.s~\ref{Figure 5}-\ref{Figure 6} when it is illuminated with a 1064 nm source. Charge distribution shows a dipole-like behavior in the body of the 12 nm AuNP. Units are arbitrary.  } 
\end{center}
\end{figure}

In Fig.~\ref{Figure 6}a, we depict the charge distribution for the eigen-mode of eigen-frequency $\sim$400 nm. We calculate the eigen-mode of the configuration given in Fig.~\ref{Figure 5}. A quadrupole plasmon mode can be easily identified on the 12 nm AuNP, on which EYFP molecules are attached~\cite{PS2}. Electric field of this high-energy eigenmode is also larger at the upper part of the 12 nm AuNP (see the light green-blue arrow on top of the AuNP), where an EYFP molecule is possible to be present. In Fig.~\ref{Figure 7}, we present the charge distribution for the 1064 nm illumination of the same configuration. Taking a glance at the charge distribution one can observe that a dipolar field is confined to the body of the 12 nm AuNP.

The picture in Fig.s~\ref{Figure 6} and \ref{Figure 7} well-fits to the model we describe below. The 1064 nm cw laser excites the dipole-like modes which are confined to the 12 nm AuNP. Two 1064 nm wavelength oscillations combine on the AuNP, via convective acceleration and the quantum electron pressure of the free-electron gas~\cite{Zayats2014Hydrodynamic_model}, to yield 532 nm oscillations. The overlap integral for the SH process, Eq.~(\ref{chi2}), is negligible if the conversion takes place into a plasmon mode which has an odd behavior on the bodies of the AuNPs~\cite{PS3}. Eq.~(\ref{chi2}) reveals that using only dipole-like plasmon excitations at the hot-spots (between the AuNPs), too, one can obtain only a weak $\chi^{(2)}$ strength for the SH conversion. Even though the 110 nm and 60 nm AuNPs have different dimensions, the charge distributions at the two sides are similar near the hot-spot, on the grounds of image charge. In addition, $\chi^{(2)}$ integral does not vanish on the 12 nm AuNP and it is stronger than the ones near the hot-spots. Therefore, 532 nm oscillations are expected to be generated mainly on the 12 nm AuNP~\cite{PS4}. The EYFP molecule attached on the 12 nm AuNP interacts strongly with the 53 nm oscillations (field in Fig.~\ref{Figure 6}b). This leads to the Fano resonance in the high-energy plasmon mode. In the following, we show that this interaction results in a $\sim$1000 enhancement factor in the nonlinear response due to the path interference effects.

\subsection{Demonstration of the path interference effects} \label{sec:path_inter}

Ordinarily, SHG process would be in resonance if the second LSP mode ($\hat{a}_2$) could be tuned to $\lambda_2$ = 1064/2 = 532 nm. However we show that presence of EYFP molecules can modify the spectrum of the SHG process taking place at the MNPs, due to the cancellation in the denominator of the nonlinear response. 

Including the diagonal ($\hat{a}_i^{\dagger}\hat{a}_i$) and AuNP-molecule interaction ($\hat{\rho}_{eg}\hat{a}_i$) terms to the Hamiltonian in Eq. (1), one obtains the equations of motion~\cite{Turkpence_2014}
\begin{align}
\dot{\alpha}_1&=\left(-i\omega_1-\gamma_1\right)\alpha_1-i2\chi^{(2)}\alpha_1^{*}\alpha_2 \label{EQ2}\\
&-if_1\rho_{ge}+\epsilon_pe^{-i\omega t} \nonumber\\
\dot{\alpha}_2&=\left(-i\omega_2-\gamma_2\right)\alpha_2-i\chi^{(2)}\alpha_1^2-if_2\rho_{ge} \label{EQ32}\\
\dot{\rho}_{ge}&=\left(-i\omega_{eg}-\gamma_{eg}\right)\rho_{ge}+i\left(f_1\alpha_1+f_2\alpha_2\right) \label{EQ4}\\
&\times\left(\rho_{ee}-\rho_{gg}\right)\nonumber\\ 
\dot{\rho}_{ee}&=-\gamma_{ee}\rho_{ee}+i\left[(f_1\alpha^{*}_1+f_2\alpha^{*}_2)\rho_{ge}-c.c.\right] \label{EQ5}  
\end{align}
where $\alpha_{1,2}=\left\langle \hat{a}_{1,2}\right\rangle$ describe the LSP field, $\rho_{ee} \; (\rho_{ge})$ are diagonal (off-diagonal) matrix elements of (EYFP) molecule with $e$ and $g$ referring to the excited and ground states. The 1064 nm laser field drives the first plasmon mode with frequency $\omega$, e.g. $\epsilon_pe^{-i\omega t}$. Damping rates of the LSP models are about $\gamma_1, \gamma_2\sim 0.1$ PHz, decay rate of molecule is about $\gamma_{ee}=2\gamma_{eg}\sim 0.1$ THz and $f_{1,2}$ are the strengths for the coupling of the EYFP molecule to $\hat{a}_{1,2}$ plasmon modes of the hybrid MNPs. As demonstrated in an earlier work~\cite{Turkpence_2014} time evolution simulations of Eqs.~(\ref{EQ2}-\ref{EQ5}) yield the steady state solutions of the form
\begin{align}
\alpha_1=\tilde{\alpha}_1e^{-i\omega t},\alpha_2= \tilde{\alpha}_2e^{-i2\omega t} \label{EQ6}\\
\rho_{eg}=\tilde{\rho}_{eg}e^{-i2\omega t},\rho_{ee}=\tilde{\rho}_{ee} \label{EQ7}
\end{align}
in the long time, where the variables with tilde are the steady-state values. $|\tilde{\alpha}_{1,2}|^2$ gives the number of plasmons oscillating with the frequency $\omega$ ($2\omega$) in the $\hat{a}_{1,2}$ LSP mode. 

In order to be able to demonstrate the physics (path interference) underlying in the enhancement/suppression scheme using analytical expressions; we neglect the coupling of $\hat{a}_1$ mode to the quantum emitter, i.e. $f_1=$0. Using solutions (\ref{EQ6})-(\ref{EQ7}) in Eq.s~(\ref{EQ2})-(\ref{EQ5}), one obtains the simple equation~\cite{Turkpence_2014} 
\begin{equation}
\tilde{\alpha}_2=\frac{i\chi^{(2)}}{\frac{|f_2|^2 y}{i(\omega_{eg}-2\omega)+\gamma_{eg}}-[i(\omega_2-2\omega)+\gamma_2]}\tilde{\alpha}^2_1 \label{EQ8}
\end{equation}
for the steady state, where $y\simeq -1$ is the population inversion ($\rho_{ee}-\rho_{gg}$) of the EYFP molecule~\cite{PS5}. Both path interference effects can be simply read from Eq. (\ref{EQ8}). Since $\lambda_2 \neq 532$ nm, the off-resonant $i(\omega_2-2\omega)$ term in the denominator reduces the SHG. However, we can arrange the imaginary part of the first term in the denominator such that, its imaginary part cancels the $i(\omega_2-2\omega)$ term. When $\omega_{eg}$ is chosen to satisfy
\begin{equation}
|f_2|^2 y(\omega_{eg}-2\omega)+(\omega_2-2\omega)\left[(\omega_{eg}-2\omega)^2+\gamma_{eg}^2\right]=0, \label{EQ9}
\end{equation}
the $\tilde{\alpha}_2$ generation can be brought almost to the resonance
\begin{equation}
\tilde{\alpha}_2=-i\chi^{(2)}\tilde{\alpha}^2_1/\gamma_2 \: , \label{EQ10}
\end{equation}
that is the value for the resonant SH converter, e.g. {\it as if} $\lambda_2=532$ nm. Therefore, cancellation in the denominator can make an inefficient converter ($\lambda_2=400$ nm) work at resonance.

12 nm AuNP is decorated with more than one EYFP molecules. Two molecules can come very close to each other for some possible configurations. In Ref.~\cite{TasginPRB2016}, we show that presence of an extra molecule can also help to (partially) cancel the $\gamma_2$ term in the denominator which cannot be performed using a single molecule. However, this complicated situation is out of the scope of this paper.

On the contrary, when we choose $\omega_{eg}=2\omega$, SHG process can be seriously suppressed. The first term in the denominator $|f_2|^2y/\gamma_{eg}$ becomes very large, since $\gamma_{eg}$ belongs to a quantum particle and is very small compared to the other frequencies. In this case, the $[i(\omega_2-2\omega)+\gamma_2]$ term becomes negligible compared to the first one. Because of the large value of the denominator SHG is suppressed. 

\subsection{Enhancement of the SHG}

\begin{figure}[!t]
\begin{center}
\includegraphics[width=3.2in]{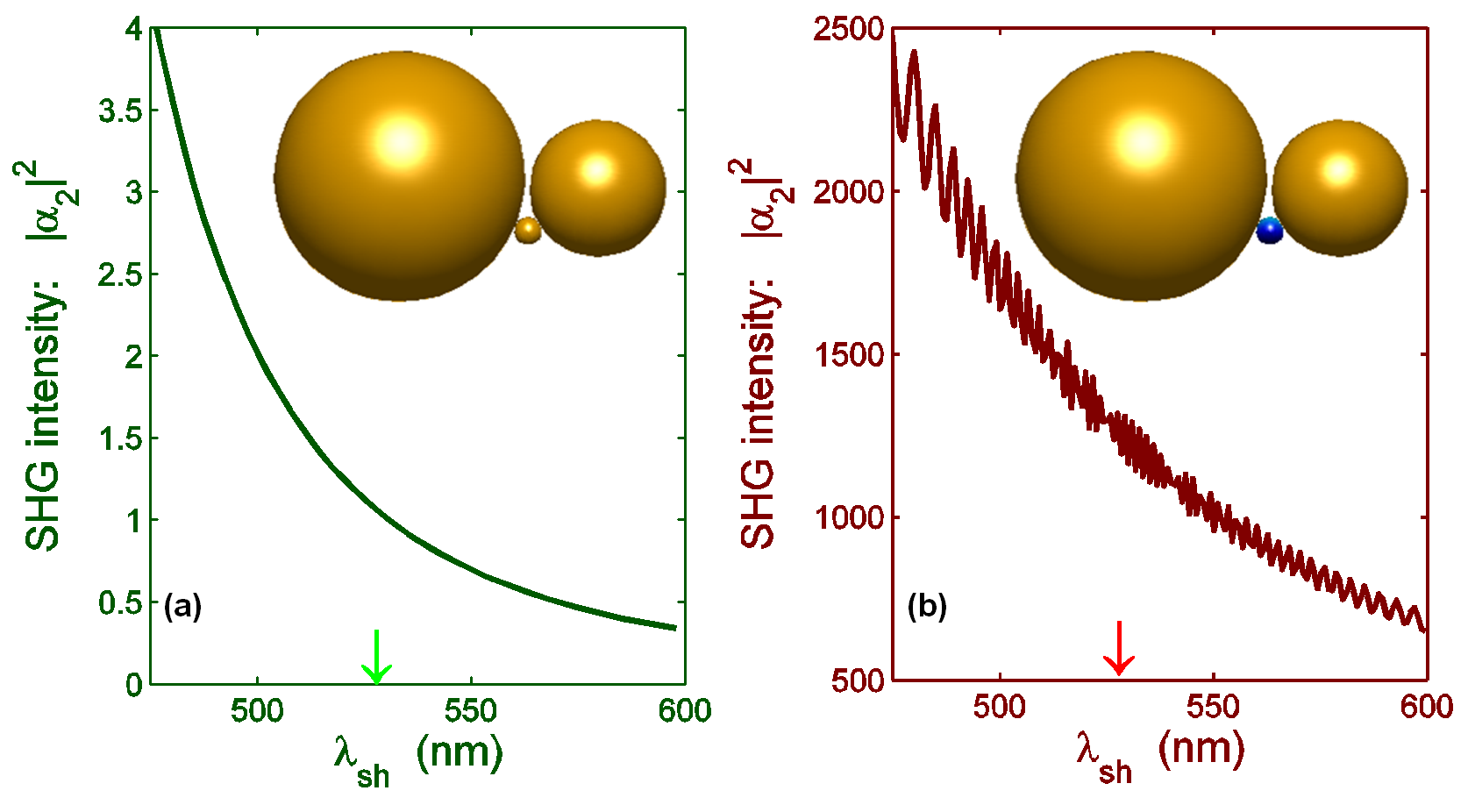}
\caption{\label{Figure 8} (color online) Enhancement of the SHG in AuNPs by insertion of an EYFP molecule. The path interference effects (See Eq. 8) due to Fano resonances can bring the nonlinear frequency conversion to resonance. For the choice of EYFP, with $\lambda_{eg}$=514 nm, SHG is enhanced $\sim$1000 times (b) compared to the no-molecule case (a). (a) Three AuNPs (of sizes 110 nm, 70 nm and 12 nm) interact. (b) The small AuNP (blue) is decorated with EYFP molecule and interacts with larger AuNPs. Arrows indicate the position of the frequency (532 nm) where SH conversion takes place in our experiment. [The parameters used in the simulation are $f_1=f_2=0.03\omega$, $\chi^{(2)}=10^{-3}\omega$, $\gamma_{1,2}=0.1\omega$, $\omega_1=1.83\omega$ ($\lambda_1=$580 nm), $\omega_2=2.66\omega$ ($\lambda_2=$400 nm), and $\omega_{eg}=2.07\omega$ ($\lambda_1=$514 nm), where frequencies are all scaled with $\omega=c/\lambda$ with $\lambda=$1064 nm. Since we present the relative enhancement in the SH field, the choice of the value of $\chi^{(2)}$ does not change these values.]} 
\end{center}
\end{figure}

In Fig.~\ref{Figure 8}, we show that presence of a molecule of level spacing $\omega_{eg}\cong$514 nm (for EYFP), SHG is enhanced about 1000 times with respect to the no-molecule case. Simulations, depicted in Fig.~\ref{Figure 8}, are obtained by time evolving Eqs. (\ref{EQ2})-(\ref{EQ5}) without neglecting the coupling of $\hat{a}_1$ mode to the quantum emitter. In deriving Eq. (\ref{EQ8}), we make such a negligence merely since we are able to demonstrate the act of path interferences.

The combined effects, (i) improved linear cross-section at 1064 nm due to hybridization, (ii) conversion enhancement due to Fano resonance, and (iii) emergence of a quadrupole-like mode on the 12 nm AuNP when placed between larger size AuNPs, make SHG observable in our experiment even when using a cw laser. We underline that unlike other studies~\cite{localization_enhanced_SHG,surface_enhanced_SHG2}, incident field (1064 nm) can be enhanced only a single order for the dimensions of the AuNPs which we use in the experiments. We also note that enhancement factor originating from the Fano resonance is to be multiplied by the enhancement due to the field localization (which is a small factor in our case).

Plasmonic nonlinear conversion of light can be a feasible route for achieving single/few molecule activation due to its strong localization and can take place with extremely high efficiency depending on the careful choice of plasmonic-molecular system construction. Simple building blocks for such systems can be localized surface plasmon (LSP) polariton supporting metal nanoparticles and their hybrids with fluorescent molecules. 

SHG takes the form of emerging oscillations of higher order (quadrupole and higher) plasmon modes. The multipolar nature of the generated SH polarization results in strong confinement of this field to the near-field and poor coupling to the far-field. When a fluorescent molecule that has absorption band in the SH frequency of the drive field is placed in the vicinity of the plasmonic system, it can efficiently absorb SH field and reradiate due to its finite electronic lifetime in the form of a dipole radiation, and hence couple effectively to the far-field. So such a fluorescent molecule can be used as a reporter of SHG.

\subsection{Controlling the Fano resonance}

Adopting the following mechanisms, it is also possible to control the value of the enhancement factor originating from the path interference effects. When the fluorescent molecules are attached on the nanoparticles through molecules with isomerism (e.g. azobenzene), the separation between the AuNP and the fluorescent molecule can be arranged with an optical control pulse. As an example, the cis and trans isomers of azobenzene have different lengths. The length of the Azobenzene linker can be controlled by illuminating with a 300-400 nm UV light. This example provides the fine tuning of the NP-molecule separation which changes the $f_2$ parameter slightly in Eq.s~(\ref{EQ2})-(\ref{EQ5}). In another method, named as DNA origami~\cite{Lazzarino_DNA_origami_2015}, hybridization of the DNA linker provides longer tuning for the NP-molecule separation, which changes $f_2$ on a larger scale.

\section{Up-conversion Mechanism} {\label{sec:upconversion}}

Throughout the paper we explain the observed up-converted signal as originating from the enhanced (Fano) SHG of AuNPs. Molecules fluoresce the $2\omega$ plasmons to the far-field as photons. One naturally suspects if the 2-photon absorption of EYFP molecules can be responsible for the signal. This possibility is excluded after the following observations. 

First, the 1064 nm field is only weakly enhanced near the hot-spot, see Fig.~\ref{Figure 5}, compared to the bare EYFPs. Hence, localization enhancement of the 2-photon absorption (via molecule) is almost negligible. Second, one may question if the EYFP-AuNP coupling modifies the spectrum. Comparing the 1-photon florescence curves for EYFP in solution~\cite{fluorescenceEYFP} and EYFP on AuNPs; one can observe that EYFP-AuNP spectrum shifts to higher energies by 4 nm. So, if this 4 nm shift does not originate from the differences in the calibration of two experiments~\cite{PS6} or the solution effect, 2-photon absorption should be less favored compared to the bare EYFPs. Because, the final energy level (the 2 photons excite) of the molecule shifts up~\cite{PS7}. 

One may still argue if modification of the virtual levels (mid-levels) of EYFP for $\sim$1064 nm energies can help the 2-photon absorption of EYFP. This, however, is quite infeasible. Because, EYFP-AuNP coupling, at hot-spots, have the electromagnetic nature and it is weak for small field enhancements at $\sim$ 1064 nm wavelengths. 

Additionally, if we attributed the up-converted signal to the 2-photon absorption of EYFP, we needed to put $\sim$1000 times Fano enhancement aside. Because, coupling of a larger band-width object (AuNP here) to a sharp band-width SH converter (molecule) does not enhance the SHG of the converter via path interference.

%We would like to point that we do not observe the SHG signal directly, but the fluorescence of the EYFP molecule due to strong absorption of the SHG signal by the EYFP. In our interpretation this is due to the fact that the EYFP molecules are well within the Förster radius of the plasmonic AuNPs resulting in a highly effective F\"{o}rster resonant energy transfer (FRET) of the SHG polarization to the molecule. 

\section{Summary and Conclusions} {\label{con}}

In summary, we experimentally show that it is possible to excite few EYFP molecules located at the gaps of asymmetric AuNP aggregates by using a cw laser excitation source. We observe the fluorescence of EYFP(s) centered at 529 nm when the aggregate is illuminated by a 1064 nm cw laser source. The two clusters, one containing AuNPs with dispersed sizes 50-120 nm and the other one containing 12 nm AuNPs decorated with EYFP molecules, do not radiate a high-frequency (i.e. 529 nm or 532 nm) signal. When we bring the two clusters together on a surface (Fig.~\ref{Figure 1}d), 12 nm AuNP enters in between the two large AuNPs. This makes the induction of a quadrupole-like plasmon mode, see Fig.~\ref{Figure 6}, on the 12 nm AuNP possible. This is the mode in which SH conversion (532 nm oscillations) can take place due to the selection rules for the SHG process~\cite{Neacsu_2005,Zayats_overlap}.

The EYFP molecules serve for two purposes. They are attached on the 12 nm AuNPs, where SH plasmon oscillations are allowed to emerge~\cite{Grosse_2012}. EYFP(s) interact with this mode (see Fig.~\ref{Figure 6}b) and yields the path interference effect in the SH conversion process which enhances SH plasmon production about 1000 times, see Fig.~\ref{Figure 8}. EYFPs additionally serve as far-field reporters. The 532 nm oscillations emerge in the quadrupole-like plasmon mode on 12 nm AuNP. Quadrupole modes couple far-field weakly. Since 532 nm plasmons are already small in number, we cannot observe the 532 nm signal on top of the 529 nm - centered fluorescence signal. EYFPs absorb the 532 nm plasmons and fluoresce to far-field much strongly compared to the coupling of the quadrupole-like plasmons.

Research on nonlinear behavior of plasmon excitations is finding its niche in diverse applications ranging quantum optical applications such as single-photon switches and single-photon transistors owing to the fact that coupling of plasmon excitation to low dimensional materials with quantum mechanical properties, such as graphene, yields pronounced nonlinear behavior~\cite{Gullans_2013} with strong plasmon field enhancement~\cite{Thongratt_2013}, as well as molecular spectroscopy such as Coherent anti-Stokes Raman scattering using Fano resonances to yield single-molecule sensitivity~\cite{Zhang_2014}. In a recent work, we show that the pronounced plasmonic activity can further be boosted by coupling of two or more quantum oscillators to plasmonic resonators, interestingly in an unlimited fashion (in principle to the point of divergence), when the strengths of inter-particle interactions and the energy level spacing for quantum oscillators are chosen carefully~\cite{Tasgin_2014}.

\begin{acknowledgments}
The research leading to these results has received funding from:  Middle East Technical University BAP-08-11-2011-129 grant; Bilim Akademisi - The Science Academy, Turkey BAGEP 2013 award (A.B.). A.B. acknowledges support from T\"{U}B\.{I}TAK grant no 113F239. M.E.T. and D.T. acknowledge support from T\"{U}B\.{I}TAK grant nos 112T927 and 114F170. This work was undertaken while one of the authors (M.E.T.) was in residence at Bilkent University with the support provided by O\u{g}uz G\"{u}lseren.
\end{acknowledgments}

%%%%%%%%%%%%%%%%%%%%%%%%%%%%%%%%%%%%%%%%%%%%%%%%%%%%%%%%%%%%%%%%%%%


\begin{thebibliography}{99}
%BIBLIOGRAPHY---BIBLIOGRAPHY---BIBLIOGRAPHY---BIBLIOGRAPHY---BIBLIOGRAPHY---BIBLIOGRAPHY---BIBLIOGRAPHY---
%%%%%%%%%%%%%%%%%%%%%%%%%%%%%%%%%%%%%%%%%%%%%%%%%%%%%%%%%%%%%%%%%%%%%%%%%%%%%%%%%%%%%%%%%%%%%%%%%%%%%%%%%%%%

%\bibitem{termo-kitap} Y. A. Cengel and M. A. Boles,{\it Thermodynamics. An Engineering Approach} (McGraw-Hill, New York, %2001).
%\bibitem{Carnot59} H. E. D. Scovil and E. O. Schulz-DuBois, Phys. Rev. Lett. {\bf2},262 (1959). 
%\bibitem{EqCarnot} J. E. Geusic, E. O. Schulz-DuBios, and H. E. Scovil Phys. Rev. {\bf 156}, 343 (1967).
%Efficiency of coherent amplification by atomic engines
%\bibitem{EfAmplf}  G. Compagno and F. Persico Phy. Lett. A {\bf65}, 103 (1978). 
%A quantum mechanical open system as a model of a heat engine
%\bibitem{OpenKoslof}  R. Kosloff, J. Chem. Phys. {\bf80}, 1625 (1984).  
%Three-level quantum amplifier as a heat engine: A study in finite-time thermodynamics 
%\bibitem{3-LampKoslov} E. Geva and R. Kosloff, Phys. Rev. E {\bf49}, 3903 (1994).
%%%%%%%%%%%%%%%%%%%%%%%%%%%%%%%%%%%%%%%%%%%%%%%%%%%%%%%%%%%%%%%%%%%%%%%%%%%%%%%%%%%%%%%%%%
\bibitem{Browne_2009} W. R. Browne and B. L. Feringa, Annu. Rev. Phys. Chem. {\bf 60}, 407 (2009). 
\bibitem{Xiang_2012} D. Xiang, Fabrication and utilization of mechanically controllable break junction for bioelectronics. PhD dissertation, Aachen University (2012).  
\bibitem{Tao_2003} B. Xu and N. J. Tao, Science {\bf 301}, 1221 (2003).   
\bibitem{Gourdon_1998} A. Gourdon, Eur. J. Org. Chem. {\bf 1998}, 2797 (1998). 
\bibitem{Bautista_2012} G. Bautista, M. J. Huttunen, J. Mäkitalo, J. M. Kontio, J. Simonen and M. Kauranen, Nano Lett. {\bf 12}, 3207 (2012).

\bibitem{Kauranen_2012} M. Kauranen, and A. V. Zayats, Nature Photonics {\bf6}, 737 (2012).


%Surface plasmon enhanced SHG from a hemicyanine monolayer.
\bibitem{Girling} I. R. Girling, N. A. Cade, P. V. Kolinsky, G. H. Cross, and I. R. Peterson, J. Phys. D: App. Phys. {\bf 19}, 2065 (1986).

\bibitem{Palomba_2009} S. Palomba, M. Danckwerts and L. Novotny, J. Opt. A: Pure Appl. Opt. {\bf11}, 114030 (2009). 

\bibitem{Neacsu_2005} C. C. Neacsu, G. A. Reider and M. B. Raschke,  Phys. Rev. B {\bf 71}, 201402(R) (2005).

%Nonlinearly coupled localized plasmon resonances: Resonant second-harmonic generation.
\bibitem{Zayats_overlap} P. Ginzburg, A. Krasavin, Y. Sonnefraud, A. Murphy, R. J. Pollard, S. A. Maier, and A. V. Zayats, Phys. Rev. B {\bf 86}, 085422 (2012).

\bibitem{Manjavacas_2011} A. Manjavacas,  F. J. García de Abajo and P. Nordlander, Nano Lett. {\bf11}, 2318 (2011).
\bibitem{Artuso_2008} R. D. Artuso and G. W. Bryant, Nano Lett. {\bf8}, 2106 (2008).
\bibitem{Waks_2010} E. Waks and D. Sridharan, Phys. Rev. A {\bf82}, 043845 (2010).
\bibitem{Weis_2011} P. Weis, J. L. Garcia-Pomar, R. Beigang and M. Rahm, Opt. Express {\bf 19}, 23573 (2011).
\bibitem{Tasgin_2010} M. E. Ta\c{s}g{\i}n, Nanoscale {\bf5}, 8616 (2010).
\bibitem{Alzar} CL Garrido Alzar, M. A. G. Martinez, and P. Nussenzveig. "Classical analog of electromagnetically induced transparency." American Journal of Physics 70.1 (2002): 37-41.

%Low-loss metamaterials based on classical electromagnetically induced transparency
\bibitem{SoukoulisPRL2009}P.  Tassin, L. Zhang, Th Koschny, E. N. Economou, and Costas M. Soukoulis, Phys. Rev. Lett {\bf 102}, 053901 (2009).


\bibitem{Butet_2012} J. Butet et al., Phys. Rev. B {\bf86}, 075430 (2012).
\bibitem{Thyagarajan_2013} K. Thyagarajan, J. Butet and O. J. F. Martin, Nano Lett. {\bf13}, 1847 (2013).
\bibitem{Berthelot_2012} J. Berthelot et al., Opt. Express {\bf20}, 10498 (2012).
\bibitem{Walsh_2013} G. F. Walsh, and  L. D. Negro, Nano Lett. {\bf13}, 3111 (2013).
\bibitem{MultiFano2016} S.-D. Liu, E. S. P. Leong, G.-C. Li, Y. Hou, J. Deng, J. H. Teng, H. C. Ong, and D. Y. Lei, ACS Nano {\bf 10}, 1442 (2016).

\bibitem{YildizJOpt2015} B. C. Yildiz, M. E. Tasgin, M. K. Abak, S. Coskun, H. E. Unalan, A. Bek, J. Opt. {\bf 17}, 125005 (2015).

%Second-harmonic generation from metal nanoparticles: resonance enhancement versus particle geometry
\bibitem{localization_enhanced_SHG} R. Czaplicki, J. Maakitalo, R. Siikanen, H. Husu, J. Lehtolahti, M. Kuittinen, and Ma. Kauranen,  Nano Lett. {\bf 15}, 530 (2014).

%Surface-Enhanced Second-Harmonic Generation
\bibitem{surface_enhanced_SHG2} C. K. Chen, A. R. B. de Castro, and Y. R. Shen, Phys. Rev. Lett. {\bf 46}, 145 (1981).


\bibitem{Turkpence_2014} D. T\"{u}rkpen\c{c}e, G. B. Akg\"{u}\c{c}, A. Bek, and M. E. Ta\c{s}g{\i}n, J. Opt. {\bf 16}, 105009 (2014).

%Enhancement of the second-harmonic generation in a quantum dot–metallic nanoparticle hybrid system
\bibitem{MahiSingh2013} M. R. Singh, Nanotechnology {\bf 24}, 125701 (2013).
\bibitem{Tasgin_2014} M. E. Ta\c{s}g{\i}n, Divergent nonlinear optical response of three resonator system via Fano resonances. Preprint at http://arxiv.org/abs/1404.3901 (2014).

\bibitem{TasginPRB2016} S. K. Singh, M. K. Abak, and M. E. Tasgin, Phys. Rev. B {\bf 93}, 035410 (2016).



% Plasmon-photon interaction in metal nanoparticles: second-quantization perturbative approach
\bibitem{2nd_quantized} M. Finazzi and F. Ciccacci, Phys. Rev. B {\bf 86}, 035428 (2012).

\bibitem{Hohenester_2012} U. Hohenester and A. Trugler, Comput. Phys. Commun. {\bf 183}, 370 (2012).   
\bibitem{Storhoff_1998} J. J. Storhoff, R. Elghanian, R. C Mucic, C. A. Mirkin and R. L. Letsinger, J. Am. Chem. Soc. {\bf 120}, 1959 (1998). 
\bibitem{Haiss_2007} W. Haiss, N. K. Thanh, J. Aveyard and D. Fernig, Anal. Chem. {\bf 79}, 4215 (2007).
\bibitem{Turkevich_1951} J. Turkevich, P. C. Stevenson and J. A. Hillier, Faraday Soc. {\bf 11}, 55 (1951).
\bibitem{Enüstün_1963} B. V. En\"{u}st\"{u}n and J. Turkevich, J. Am. Chem. Soc. {\bf 85}, 3317 (1963).
\bibitem{Ojea-Jiménez_2010} I. Ojea-Jiménez, F. M Romero, N. G. Bastús and V. Puntes, The J. of Phys. Chem. C {\bf 114}, 1800 (2010).
\bibitem{Kimling_2006} J. Kimling, M. Maier, B. Okenve, V. Kotaidis, H. Ballot and A. Plech, The J. of Phys. Chem. B {\bf 110}, 15700 (2006).
\bibitem{Hazarika_2006} P. Hazarika, F. Kukolka and C. M. Niemeyer,  Angew. Chem. {\bf110}, 6827 (2006).
\bibitem{Kukolka_2004} F. Kukolka and C. M. Niemeyer, Org. Biomol. Chem. {\bf 2}, 2203 (2004).
\bibitem{Niemeyer_2003} C. M. Niemeyer, B. Ceyhan and P. Hazarika, Angew. Chem. {\bf 42}, 5766 (2003).
\bibitem{Blab_2001} G. A. Blab, P. H. M. Lommerse, L. Cognet, G. S. Harms, and T. Schmidt, Chem. Phys. Lett. {\bf 350}, 71 (2001).



\bibitem{Nordlander_2004} P. Nordlander, C. Oubre, E. Prodan, K. Li and M. I. Stockman, Nano Lett. {\bf 4}, 899 (2004).
\bibitem{Günendi_2013} M. C. G\"{u}nendi, \.{I}. Tanyeli, G. B. Akg\"{u}\c{c}, A. Bek, R. Turan and O. G\"{u}lseren Opt. Express {\bf21}, 18344 (2013). 
\bibitem{Grosse_2012} N. B. Grosse, J. Heckmann and U. Woggon,  Phys. Rev. Lett. {\bf 108}, 136802 (2012).
\bibitem{Bouhelier_2005} A. Bouhelier, R. Bachelot, G. Lerondel, S. Kostcheev, P. Royer and G. P. Wiederrecht, Phys. Rev. Lett. {\bf95}, 267405 (2005).
\bibitem{Note} In our experiment, SHG is reported to far-field by fluorescent molecules.


\bibitem{Pelton_book_2013} M. Pelton and G. W. Bryant, "{\it Introduction to metal-nanoparticle plasmonics}, John Wiley \& Sons, (2013).

\bibitem{PS2} In the simulation, we considered only the plasmonic particles, the 3 AuNPs. EYFP molecules are not included in the simulation.

\bibitem{PS3} In Fig.~\ref{Figure 7}, the 1064 nm E-field on top of the 12 nm AuNP, which is generated by the charges on the surfaces of the 110 nm and 60 nm AuNPs, also helps the induction of SH conversion on the 12 nm AuNP.

%Nonperturbative hydrodynamic model for multiple harmonics generation in metallic nanostructures.
\bibitem{Zayats2014Hydrodynamic_model} P. Ginzburg, A. V. Krasavin, G. A. Wurtz, and A. V. Zayats, ACS Photonics {\bf 2}, 8 (2014).

\bibitem{PS4} At/near this resonance ($\lambda_1$=400 nm) there could be other modes, other than quadrupole profile. This would not affect the description  much. Because SHG can emerge to the quadrupole mode.

\bibitem{PS5} EYFP is only weakly excited since the generated SH field is very very weak.

\bibitem{PS6} We cannot be sure if the 4 nm shift in 1-photon florescence of EYFP emerges from EYFP-AuNP coupling, due to calibration of our experiment and setup in Ref.~\cite{fluorescenceEYFP}, or because of the presence of solution in Ref.~\cite{fluorescenceEYFP}. The two spectra appear as the shifted form of each other. We cannot observe an differentiation in the two curves indicating EYFP-AuNP hybridization. Note that, for the emergence of Fano resonance a weaker coupling, not introducing hybridization, is enough. 

\bibitem{PS7} When the total excitation energy ($2\omega$) is shifted up relative to the 2-photon absorption level, the vibrational/rotational states of the molecule can help the 2-photon process. However, when $2\omega$ is shifted down compared to the final 2-photon level, energy conservation works agains the 2-photon absorption process.

%"Two‐photon excitation and emission spectra of the green fluorescent protein variants ECFP, EGFP and EYFP
\bibitem{fluorescenceEYFP} E. Spiess, F. Bestvater, A. Heckel-Pompey, K. Toth, M. Hacker, G. Stobrawa, T. Feurer, C. Wotzlaw, U. Berchner-Pfannscmidt, T. Porwol, and H. Acker, J. Microscopy {\bf 217}, 200 (2005).


%Plasmon resonance tuning using DNA origami actuation
\bibitem{Lazzarino_DNA_origami_2015} L. Piantanida, D. Naumenko, E. Torelli, M. Marini, D. M. Bauer, L. Fruk, G. Firrao, and M. Lazzarino, Chem. Comm. {\bf 51},4789 (2015).

%%Seeded growth of submicron Au colloids with quadrupole plasmon resonance modes
%\bibitem{Abajo_quadrupole_2006} J. Rodriguez-Fernandez, J. Perez-Juste, F. J. Garcia de Abajo, and L. M. Liz-Marzan, Langmuir {\bf 22}, 7007 (2006).

\bibitem{Senet_1996} P. Senet, J. Chem. Phys. {\bf 105}, 6471 (1996).
\bibitem{Lohner_1999} F.P Lohner and A. A. Villaeys, Surf. Sci. {\bf 422}, 17 (1999).
\bibitem{Schaich_2000} W. L. Schaich, Phys. Rev. B {\bf 61}, 10478 (2000).
\bibitem{Bachelier_2010} G. Bachelier, J. Butet, I. Russier-Antoine, C. Jonin, E. Benichou and P. F. Brevet, Phys. Rev. B  {\bf 82}, 235403 (2010).
\bibitem{Liebsch_1988} A. Liebsch, Phys. Rev. Lett. {\bf61}, 1233 (1988).

\bibitem{Gullans_2013} M. Gullans, D. E. Chang, F. H. L Koppens, F. J. Garc\'ia de Abajo and M. D. Lukin, Phys. Rev. Lett. {\bf 111}, 247401 (2013).
\bibitem{Thongratt_2013} S. Thongrattanasiri and F. J. Garc\'ia de Abajo, Phys. Rev. Lett. {\bf 110}, 187401 (2013).
\bibitem{Zhang_2014} Y. Zhang, Y. R. Zhen, O. Neumann, J. K. Day, P. Nordlander and N. J. Halas, Nat. Commun. {\bf 5}, 4424 (2014).
\end{thebibliography}
\end{document}